\documentclass[universe,article,accept,pdftex,moreauthors]{Definitions/mdpi}
\usepackage{xcolor}

\firstpage{1} 
\pubvolume{1}
\issuenum{1}
\articlenumber{0}
\pubyear{2026}
\copyrightyear{2026}
\externaleditor{Firstname Lastname} 
\datereceived{15 June 2026} 
\daterevised{17 August 2026} 
\dateaccepted{21 August 2026} 
\datepublished{ } 

\Title{\textls[-15]{Exploring Late Stellar Evolution in the Era of Large Surveys: Machine Learning Prospects for Hot Subdwarfs and White Dwarfs}}

\Author{Princy Ranaivomanana $^{1}$\orcidA{} and Murat Uzundag $^{2,}$$^{*}$\orcidB{}}

\AuthorNames{Princy Ranaivomanana and Murat Uzundag}

\address{%
$^{1}$ \quad Department of Physics, University of Antananarivo, P.O. Box 906, Antananarivo 101, Madagascar;  rtprincy@gmail.com\\
$^{2}$ \quad Instituut voor Sterrenkunde, KU Leuven, Celestijnenlaan 200D, 3001 Leuven, Belgium}

\corres{Correspondence: muratuzundag.astro@gmail.com}

\abstract{The rapid growth of large-scale astronomical surveys and advances in data-driven analysis techniques have transformed the study of late-stage stellar evolution. Modern facilities are producing large volumes of photometric, spectroscopic, and astrometric data, enabling systematic investigations of compact stellar populations across the Milky Way. Among the most important tracers of these advanced evolutionary phases are hot subdwarfs and white dwarfs: hot subdwarfs are core-helium-burning tracers of late, binary-driven stellar evolution, while white dwarfs represent the final evolutionary endpoint of low- and intermediate-mass stars. These compact objects provide important laboratories for studying stellar interiors, binary evolution, and the long-term fate of planetary systems.
This paper explores how recent advances in machine learning are being applied to the detection, characterization, and, when combined with follow-up spectroscopy and modeling, the physical interpretation of hot subdwarfs and white dwarfs. By combining photometric, spectroscopic, and time-domain observations with these computational tools, it is now possible to efficiently discover rare objects, detect stellar variability, and probe the internal structure and evolutionary pathways of compact stars. Ultimately, these developments highlight the growing role of advanced algorithms in supporting the study of the final stages of stellar evolution, provided their outputs are validated against physical observables.}


\newcommand{\apj}{Astrophys. J.}
\newcommand{\apjs}{Astrophys. J. Suppl. Ser.}
\newcommand{\apjl}{Astrophys. J.}
\newcommand{\aap}{Astron. Astrophys.}

\newcommand{\aj}{Astron. J.}
\newcommand{\mnras}{Mon. Not. R. Astron. Soc.}
\newcommand{\araa}{Annu. Rev. Astron. Astrophys.}

\newcommand{\nar}{New Astron. Rev.}

\newcommand{\pasp}{Publ. Astron. Soc. Pac.}
\begin{document}


\section{Introduction}
\label{intro}

Stars are the fundamental building blocks of galaxies and play a central role in driving the physical and chemical evolution of the Universe. Through their life cycles, stars synthesize heavy elements, regulate the energetics of the interstellar medium, and ultimately shape the formation and evolution of planetary systems. Although the general framework of stellar evolution is well established, important uncertainties remain in our understanding of the internal physical processes that govern stellar structure and evolution. Key mechanisms, including rotation, internal mixing, magnetic fields, and chemical stratification, are particularly difficult to constrain observationally, especially during the late stages of stellar evolution when stars evolve rapidly toward compact remnants~\citep{2021RvMP...93a5001A}.

Over the past decade, astronomy has entered an unprecedented era of data-rich discovery driven by large photometric, spectroscopic, and astrometric surveys. Space missions such as Gaia~\citep{2016aanda...595A...1G} have revolutionized stellar astrophysics by providing precise distances, photometry, and astrometry for more than a billion stars in the Milky Way. At the same time, ground-based spectroscopic programs including the Large Sky Area Multi-Object Fiber Spectroscopic Telescope (LAMOST)~\citep{2012RAA....12.1197C}, currently one of the largest spectroscopic surveys of stars, Sloan Digital Sky Survey-V (SDSS-V)~\citep{2017arXiv171103234K}, the William Herschel Telescope Enhanced Area Velocity Explorer (WEAVE)~\cite{2024MNRAS.530.2688J} and the 4-metre Multi-Object Spectroscopic Telescope (4MOST)~\citep{2019Msngr.175....3D} are delivering large-scale stellar spectroscopy across wide areas of the sky. In addition, the Legacy Survey of Space and Time, now underway at the Vera C. Rubin Observatory, is expected to dramatically expand the discovery space for variable and transient phenomena over the coming decade~\citep{2019ApJ...873..111I}. 
Together, these facilities are generating massive and heterogeneous datasets that enable systematic studies of stellar populations across the Galaxy, but also require new computational and statistical methods to extract meaningful physical insights.

Among the most important objects for understanding the final stages of stellar evolution are hot subdwarfs and white dwarfs. Hot subdwarfs (sdO/B stars) are low-mass, core-helium-burning stars located at the extreme blue end of the horizontal branch in the Hertzsprung--Russell (HR) diagram~\citep{2009ARA&A..47..211H, 2016PASP..128h2001H,2024arXiv241011663H}. Their positions in an observational HR diagram (or color-magnitude diagram) are illustrated in Figure~\ref{fig:hrd}, using Gaia DR3 data for all objects within 1 kpc (see Appendix A.1 in~\citep{2025A&A...704A..70R} for data query details). They are widely believed to form primarily through binary evolution channels involving significant mass loss, such as stable Roche-lobe overflow, common-envelope ejection, or the merger of helium white dwarfs, although single-star formation channels (e.g., via enhanced mass loss on the red giant branch) have not been completely excluded and may account for a minority of the population~\citep{2016PASP..128h2001H}. These stars are also thought to be responsible for the ultraviolet excess observed in elliptical galaxies and provide important constraints on mass-loss processes during the red giant phase~\citep{2013A&ARv..21...59I,2003MNRAS.341..669H, 2002MNRAS.336..449H}. A subset of hot subdwarfs exhibits pulsations~\citep{1997ApJ...483L.123C,1997MNRAS.285..640K,2003ApJ...583L..31G} that enable detailed probing of their interiors through asteroseismology, offering unique insights into stellar interiors and evolutionary pathways~\citep{uzundag2021, 2024A&A...684A.118U}.

White dwarfs represent the final evolutionary stage of the vast majority of stars, including most stars with initial masses below roughly eight solar masses, accounting for more than 95\% of the stellar population in the Milky Way~\citep{2010A&ARv..18..471A}. As compact stellar remnants supported by electron degeneracy pressure, white dwarfs preserve valuable information about the evolutionary histories of their progenitor stars. Many white dwarfs exhibit photometric variability caused by pulsations, binarity, rotation, or the presence of orbiting debris from disrupted planetary bodies. Pulsating white dwarfs in particular serve as powerful laboratories for asteroseismology, enabling detailed investigations of their internal composition, thermal structure, and evolutionary cooling rates~\citep{WK2008,2008PASP..120.1043F,Corsico19}.

\textls[-15]{Although hot subdwarfs and white dwarfs differ substantially in evolutionary stage, formation channel, and observational selection effects, they are considered jointly in this work because they share a common set of data-analysis challenges: both are intrinsically rare compared to main-sequence contaminants and are easily confused with other source classes in photometric surveys. In addition, hot subdwarfs and white dwarfs increasingly rely on the same family of machine learning (ML) tools (automated candidate selection, variability classification, and parameter inference) to be efficiently identified and characterized in current and forthcoming large-scale surveys. Hot subdwarfs additionally represent a relevant evolutionary precursor to a fraction of white dwarfs formed through binary mass-loss channels, further motivating a combined discussion of their machine-learning-driven study.}

\textls[-15]{Despite significant progress in stellar astrophysics, several fundamental questions remain unresolved. The formation channels of hot subdwarfs and the role of binary interactions in shaping their evolution remain active areas of investigation. Similarly, the interpretation of pulsation spectra in pulsating compact stars requires detailed theoretical modeling to connect observed frequencies with internal structural properties. In addition, the discovery of planetary debris disks and metal-polluted atmospheres around many white dwarfs has opened a new window into the long-term fate of planetary systems after the death of their host stars. These systems provide valuable opportunities to study the composition and dynamical evolution of planetary material long after the main-sequence phase~\citep{2016RSOS....350571V,2014MNRAS.445.2244V}.}

The large volume and diversity of data produced by modern astronomical surveys are reshaping the study of compact stellar remnants. Large-scale datasets from missions such as Gaia, combined with continuous observations from space-based photometric surveys, such as Kepler/K2~\citep{borucki2010, haas2014} and Transiting Exoplanet Survey Satellite (TESS)~\citep{Ricker2015}, are enabling the discovery and characterization of large populations of hot subdwarfs and white dwarfs. At the same time, the scale and complexity of these datasets necessitate the use of data-driven methodologies, including ML techniques, automated classification pipelines, and advanced statistical analysis~\citep{2025A&A...693A.268R}. In this context, data-driven approaches are emerging as effective tools for identifying rare stellar populations, classifying variable stars, and extracting subtle astrophysical signals from large survey datasets. By combining large photometric and spectroscopic surveys with modern computational methods, it is now possible to investigate the late stages of stellar evolution with improved statistical power and precision. In this review, we examine how ML and data-driven analysis techniques are being applied to large astronomical surveys to advance our understanding of hot subdwarfs, white dwarfs, and the complex physical processes that govern the final stages of stellar evolution.

\begin{figure}[H]
\includegraphics[width=0.75\linewidth]{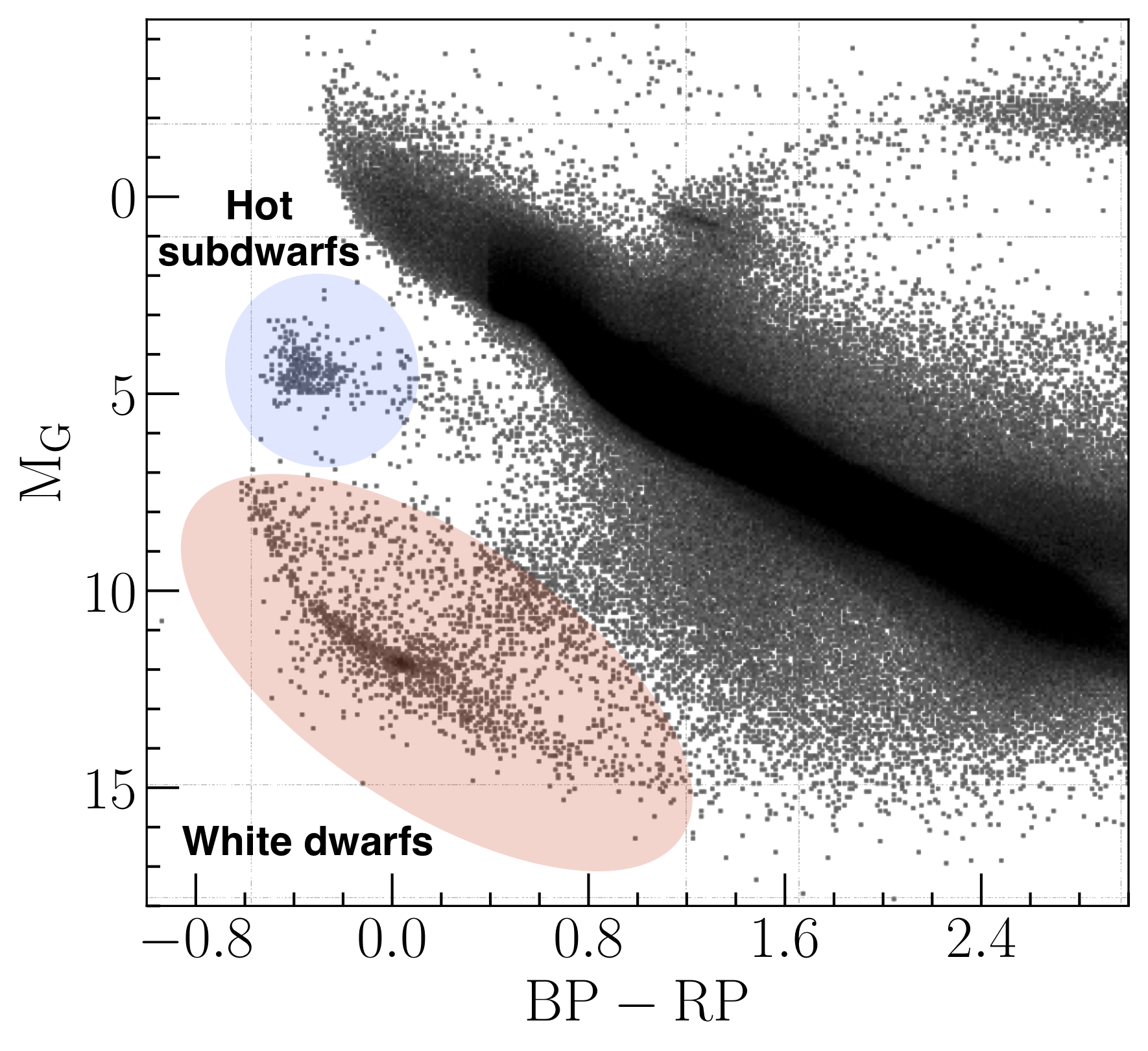}
\caption{Observational HR diagram depicting hot subdwarfs (blue-shaded area) and white dwarfs (red-shaded area) from the Gaia DR3 catalog. The sample is restricted to sources within 1~kpc with reliable parallaxes (relative parallax uncertainty $<$20\%) and a re-normalized unit weight error (RUWE) $<$1.4, and no extinction correction has been applied. The blue- and red-shaded regions are illustrative loci intended to guide the eye to the approximate location of hot subdwarfs and white dwarfs, respectively, rather than formal selection boundaries; see Appendix A.1 of~\citep{2025A&A...704A..70R} for the full query and selection criteria. Note that this 1~kpc distance cut is the reason the plotted white dwarf sample appears comparatively small: the full Gaia DR3 white dwarf catalog contains over 360,000 candidates~\citep{2021MNRAS.508.3877G}, and the population shown here is not intended to be representative of that full sample size, but only of the region it occupies in the HR diagram.}
\label{fig:hrd}
\end{figure}

We stress that this article is positioned as a narrative, topical review rather than a systematic or fully quantitative meta-analysis: it does not introduce a new algorithm, benchmark dataset, or quantitative comparison of methods, but instead synthesizes recent applications of ML to hot subdwarfs and white dwarfs, following an explicit (though not exhaustive) selection of the literature described below, and identifies the key methodological gaps and future directions that we consider most relevant for the community.

\subsection*{Scope and Selection of Literature}

Given the breadth of both the ML literature and the surveys relevant to late stellar evolution, we restrict the scope of this review as follows. In terms of ML tasks, we focus on candidate identification, variable-source and spectral classification, binary detection, atmospheric-parameter inference, and unsupervised discovery of rare subclasses, since these represent the tasks most extensively addressed for hot subdwarfs and white dwarfs to date. In terms of surveys, we emphasize Gaia, TESS, ZTF, SDSS, LAMOST, and the forthcoming Rubin/LSST, WEAVE, and 4MOST facilities, as these dominate the current and near-future data volume relevant to compact stellar remnants. We selected hot subdwarfs and white dwarfs, rather than other compact or evolved stellar populations, as representative case studies of late stellar evolution because they are the two classes for which ML applications have matured most rapidly over the past five years, span the full range of supervised and unsupervised techniques discussed in Section~\ref{methods}, and illustrate, in a directly comparable way, the challenges of rare-object identification in large, contaminated survey datasets. The literature reviewed here was primarily selected from works published over the last decade, with an emphasis on the last five years, and does not aim at an exhaustive literature coverage.

The paper is organized as follows: Section~\ref{methods} provides an overview of data-driven methods and ML techniques commonly used in stellar astrophysics, with an emphasis on their application to large and heterogeneous survey data. Section~\ref{methods_for_sdbs} focuses on the application of these approaches to hot subdwarfs, including their identification, variability classification, and characterization in the context of large photometric and spectroscopic datasets. Section~\ref{methods_for_wds} presents similar methodologies applied to white dwarfs, highlighting advances in variability detection, spectral analysis, and the study of planetary debris and binary systems. Finally, Section~\ref{conclusion} summarizes the main conclusions and findings from the literature and discusses future prospects for the study of late stellar evolution in the era of next-generation astronomical surveys and data-driven discovery.

\section{Data-Driven Methods in Stellar Astrophysics}
\label{methods}
The integration of ML techniques into stellar astronomy is a rapidly evolving domain. This is because astronomy has entered the big data era, with facilities such as Vera Rubin producing tens of terabytes nightly, while astrometric and spectroscopic instruments such as Gaia, WEAVE, and 4MOST generate complex downstream catalogs (see Table\,~\ref{tab:data_rates_facilities}). Scalable, data-driven solutions are therefore required to leverage the wealth of data generated by these large surveys. The past decade has seen the rapid development of a broad spectrum of ML architectures in stellar astronomy, ranging from classical algorithms to advanced deep learning models. These algorithms were specifically designed to address common challenges driven by the large volume of data produced by modern astronomy facilities. These challenges are summarized in Table~\ref{tab:ml_challenges_facilities} for each facility, while standard ML methods applied in stellar astronomy are reviewed in the following sections.

\begin{table}[H]
\small
\caption{Approximate data volume and acquisition rates for six modern optical astronomy facilities dedicated to stellar science. Values represent typical raw science data or operational averages and may vary with observing mode. The ``data volume'' column reports the processed or downlinked science product per day/night, while the ``instrumental rate'' column reports the underlying camera readout or downlink rate; these two quantities are of a different physical nature (integrated product vs. instantaneous rate) and should not be compared directly across facilities.}
\label{tab:data_rates_facilities}
\begin{adjustwidth}{-\extralength}{0cm}
\begin{tabularx}{\fulllength}{lLLll}
\toprule
\textbf{Facility} & \textbf{Stellar Science Role} & \textbf{Data Volume} & \textbf{Instrumental Rate} & \textbf{Ref.}\\
\midrule

Gaia & 
Astrometry, photometry, spectroscopy &
$\sim$40 GB day$^{-1}$ (downlink) &
1--8.5 Mbps (telemetry) &
\cite{2016aanda...595A...1G}
\\

TESS &
High-cadence stellar photometry &
$\sim$14 GB day$^{-1}$ (downlink) &
$\sim$100 Mbps (telemetry) &
\cite{Ricker2015}
\\

ZTF &
Time-domain stellar variability &
$\sim$1.0--1.4 TB night$^{-1}$ (science product) &
$\sim$150 MB s$^{-1}$ (readout) &
\cite{Bellm2019,2019PASP..131g8001G}
\\

LSST &
Large-scale time-domain survey &
$\sim$15 TB night$^{-1}$ (science product) &
$\sim$3.2 GB s$^{-1}$ (readout) &
\cite{2019ApJ...873..111I,2017ASPC..512..279J}
\\

WEAVE &
Multi-object stellar spectroscopy &
15--25 GB night$^{-1}$ (science product) &
2--5 MB s$^{-1}$ (readout) &
\cite{2012SPIE.8446E..0PD,2024MNRAS.530.2688J}
\\

4MOST &
Galactic stellar spectroscopy &
20--30 GB night$^{-1}$ (science product) &
3--6 MB s$^{-1}$ (readout) &
\cite{2019Msngr.175....3D,2012SPIE.8446E..0TD}
\\

\bottomrule
\end{tabularx}
\end{adjustwidth}
\end{table}
\unskip

\begin{table}[H]
\small
\caption{Representative machine learning challenges associated with major modern stellar astronomy facilities.}
\label{tab:ml_challenges_facilities}
\begin{adjustwidth}{-\extralength}{0cm}
\begin{tabularx}{\fulllength}{llLl}
\toprule
\textbf{Facility} & \textbf{Dominant Data Type} & \textbf{ML Challenge} & \textbf{Ref.}\\
\midrule

Gaia &
Time series + catalog &
Anomaly detection, object classification, XP-spectra parameterization, astrometric-binary identification, and variability characterization &
\cite{Rimoldini2019,Rimoldini2023,Eyer2022}
\\

TESS &
Light curves &
Transit detection, denoising, stellar-variability classification, asteroseismic mode extraction, rotation-period recovery, and flare detection &
\cite{2015ApJ...809...77S,2019AJ....158...25Y,2022mla..confE..11M}
\\

ZTF &
Imaging + light curves &
Real-time transient classification &
\cite{2019PASP..131c8002M,2019MNRAS.489.3582D}
\\

LSST &
Massive imaging streams &
Alert filtering and event prioritization &
\cite{2018ApJS..236....9N}
\\

WEAVE &
Spectra &
Atmospheric-parameter and abundance inference &
\cite{Guiglion2020}
\\

4MOST &
Spectroscopic surveys &
Atmospheric-parameter and abundance inference &
\cite{Nepal2023}
\\

\bottomrule
\end{tabularx}
\end{adjustwidth}
\end{table}

\subsection{Supervised Learning in Stellar Astronomy}
Supervised ML methods are widely applied in stellar astronomy for classification or prediction tasks where data labels are available. Classical algorithms such as Random Forests (RF), Support Vector Machines (SVMs), and Gradient Boosting methods (e.g., XGBoost, LightGBM, CatBoost) frequently perform these tasks. Specifically, they have been applied to classify variable sources, derive stellar parameters, and predict both stellar ages and surface rotation periods~\cite{2011ApJ...733...10R, Anders2023, Rimoldini2023, Gomes2024, 2026MNRAS.546ag132M}. Furthermore, ensemble and gradient boosting methods, including RF and XGBoost, offer high interpretability through feature importance rankings and remain computationally efficient for large-scale classification tasks. However, these traditional ML methods require careful manual feature extraction from raw data to build training datasets. In time-domain astronomy, these features are typically descriptive statistics derived from time-series data. To extract these statistical parameters from light curves, researchers routinely rely on specialized pipelines, such as the Feature Extractor for Time Series (FEETS)~\cite{Cabral2018}, the Time Series Feature Extraction Library (TSFEL)~\cite{BARANDAS2020100456}, and Superphot+~\citep{deSoto_2024}.

While manual feature extraction is a statistically robust method for representing tabular data, it struggles to capture the complex, non-linear patterns hidden within unstructured data, such as astronomical images, light curves, and raw stellar spectra. In contrast, deep learning algorithms bypass this limitation by performing automatic representation learning, extracting hierarchical features directly from the raw data through deep neural network architectures. Among these algorithms, Convolutional Neural Networks (CNNs) are widely used for spectral analysis and identifying physical/chemical labels~\cite{2020MNRAS.491.2280S, Nepal2023, Guiglion2024}, recurrent Neural Networks (RNNs), such as Long Short-Term Memory (LSTM) and Gated Recurrent Units (GRUs), have been deployed for stellar parameter prediction~\citep{2018AJ....156....7H} and RR Lyrae stars metallicity estimation~\cite{2025A&A...702A.148M} using light curve data from TESS and Gaia DR3, respectively.

It is important to note that the performance of these supervised methods is frequently reported using overall classification accuracy, which can be misleading when applied to hot subdwarfs and white dwarfs. Both classes are strongly outnumbered by main-sequence and giant contaminants in typical survey volumes, so a classifier can achieve high accuracy while still missing a large fraction of true positives or, conversely, admitting substantial contamination. Class-imbalance-aware metrics, such as precision, recall, the F1 score, or the area under the precision--recall curve, are therefore preferable to accuracy alone when these methods are applied to rare compact-object searches. Similarly, the choice of training set itself can propagate biases: models trained on spectroscopically confirmed subsamples (e.g., from SDSS or LAMOST) may not generalize to the full photometric candidate population probed by Gaia, WEAVE, or 4MOST.

\subsection{Unsupervised Learning in Stellar Astronomy}
\textls[-15]{Supervised ML algorithms require extensive collections of accurately labeled data, such as expert-labeled light curves, to achieve adequate performance. However, generating these representative training sets is a highly resource-intensive, manual, and time-consuming effort, which is a significant bottleneck given that modern surveys produce unlabeled data at a much faster rate. To this end, unsupervised ML techniques have become essential in astronomy, as they can capture hidden structures and patterns without relying on labeled datasets. Additionally, with unsupervised learning, it is possible to discover new classes of objects or anomalies in the datasets, which is not possible with supervised learning. }

Unsupervised learning methods often apply dimensionality reduction algorithms to find meaningful representations of the data in the lower-dimensional feature space, followed by clustering algorithms to identify potential clusters. Dimensionality reduction techniques, such as t-distributed stochastic neighbor embedding (t-SNE)~\citep{Vandermaaten2008}, uniform manifold approximation and projection (UMAP)~\cite{McInnes2018}, and principal component analysis (PCA)~\cite{Jolliffe2002} have been used to cluster variable stars and discover anomalies in massive datasets~\cite{2025A&A...693A.268R, 2025A&A...704A..70R}. UMAP and t-SNE are both non-linear techniques that map high-dimensional data into a two- or three-dimensional space while trying to preserve the local neighborhood structure of the data (i.e., points that are similar in the original feature space remain close together in the projection). The two methods differ in how they build this projection: t-SNE converts pairwise distances between points into similarity probabilities, using a Gaussian distribution in the high-dimensional space and a Student's t-distribution in the low-dimensional space, and then minimizes the divergence between the two probability distributions, whereas UMAP relies on a graph-based construction rooted in topological data analysis to preserve the neighborhood structure of the data. Neither method is universally superior to the other, and their relative performance depends on the dataset and the specific clustering task at hand; in practice, it is therefore advisable to apply both algorithms to a given dataset and adopt whichever yields the more informative or better-separated embedding. Clustering algorithms, including density-based spatial clustering of applications with noise (DBSCAN)~\citep{Ester1996} and Gaussian Mixture Models (GMMs)~\citep{Dempster1977,Reynolds2009}, are in turn used to search for coherent stellar clusters and group members from astrometric parameters~\cite{2020A&A...635A..45C, 2020A&A...633A.154L, 2022A&A...664A.175P, 2023MNRAS.523.3538N}.

Caution is required when interpreting the resulting low-dimensional embeddings. Neither t-SNE nor UMAP preserves global distances in a physically meaningful way: the relative size, shape, and separation of clusters in the projected space do not necessarily correspond to genuine differences in the underlying physical parameter space and can instead be strongly affected by the choice of hyperparameters (e.g., perplexity, number of neighbors), the input feature set, and the specific sample selection used to build the embedding. Consequently, cluster boundaries identified in t-SNE or UMAP maps should be regarded as a qualitative guide for candidate selection and hypothesis generation, rather than as a quantitative or physically calibrated classification, unless independently validated against spectroscopically confirmed samples or physical models.

A specific application of this unsupervised pipeline to the clustering of hot subdwarf light curves, together with the resulting t-SNE embeddings, is presented in Section~\ref{sec3.2},
in the context of the detection of binarity and variability in this class of stars.

\subsection{Advanced Architectures}

The field is increasingly adopting advanced deep learning models, including autoencoders (standard and variational) for data compression~\cite{2025A&A...693A.256C, 2025A&A...701A.150H}, Generative Adversarial Networks (GANs)~\citep{2014Goodfellow} for disentangling physical and chemical spectral properties~\cite{2025A&A...694A.326M}, and Transformer-based models to capture long-range dependencies in light curves~\cite{2024MNRAS.528.5890P,2022mla..confE..11M,2023A&A...670A..54D,2024A&A...689A.289C}.

Advanced ML architectures, such as Transformer~\citep{2017arXiv170603762V} and Variational Autoencoders (VAEs)~\citep{2013arXiv1312.6114K}, offer transformative advantages in stellar astronomy by automatically extracting informative representations directly from raw, multimodal data, thereby eliminating the need for computationally expensive manual feature engineering~\citep{2025A&A...701A.150H}. Transformers excel at capturing long-range dependencies and broad-timescale patterns, such as stellar granulation, that traditional networks often miss, all while leveraging GPU parallelization to achieve inference speeds hundreds of times faster than classical models like Random Forests~\citep{2024MNRAS.528.5890P}. Concurrently, generative models and self-supervised autoencoders facilitate unsupervised anomaly discovery and the creation of realistic synthetic spectra, providing a vital alternative strategy when labeled observational data is scarce~\citep{2025A&A...693A.256C}. However, building and training these advanced architectures introduces significant challenges, most notably the high computational resource cost of self-supervised pre-training, which demands massive datasets and weeks of processing time before the models can even be fine-tuned for specific tasks~\citep{2023A&A...670A..54D}. Additionally, these representations can lose specific global context if their pre-training auxiliary tasks are not comprehensive enough. For generative models like VAEs, relying on standard unimodal Gaussian priors introduces statistical imprecision, making it difficult to properly model known labels with non-normal distributions or to accurately represent the inherent physical degeneracy of complex stellar spectra~\citep{2025A&A...693A.256C}.

\subsection{Machine Learning Candidates vs.
Physically Confirmed Objects}

A recurring point throughout this review is the distinction between objects flagged as candidates by an ML pipeline and objects that are physically confirmed through dedicated follow-up. For hot subdwarfs, physical confirmation typically requires spectral classification together with constraints on effective temperature, surface gravity, and helium abundance. For white dwarfs, confirming a magnetic field, metal pollution, or a binary companion similarly relies on spectroscopy, radial velocity monitoring, multi-epoch photometry, or polarimetry. We therefore regard ML primarily as a tool for discovery and priority ranking within large surveys, rather than as a final confirmation of physical classification. As such, dedicated follow-up spectroscopy and time-domain observations remain necessary to confirm the physical nature of a machine learning candidate.


\section{Applications to Hot Subdwarfs}
\label{methods_for_sdbs}

As previously mentioned, hot subdwarfs offer an ideal testbed for the study of binary stellar evolution. While their exact formation and evolution remain subjects of ongoing research, the systematic observation and identification of these stars are crucial for improving our understanding. Historically, the identification of hot subdwarf candidates relied on traditional color-cut methods followed by resource-intensive visual inspection. However, the exponential growth of multi-epoch photometric and spectroscopic data from large-scale astronomical surveys has driven the integration of ML techniques into the identification and study of these objects.

To date, most ML applications have focused on identifying hot subdwarf candidates and classifying their variability. This is primarily because hot subdwarf candidate selections in observational HR diagrams are heavily contaminated by cataclysmic variables (CVs), low-mass main-sequence stars, and white dwarfs, as shown in Figure~\ref{fig:hrd}. Consequently, robust and scalable statistical and ML methods are required to isolate hot subdwarfs from these contaminants. In addition to candidate selection, photometric variability detection has become a computationally demanding task given the large volume of time-series data generated by modern photometric and spectroscopic surveys. We categorize recent progress in ML applications to hot subdwarf studies into three primary domains: (i)
the automated identification and variability classification of candidates within massive datasets; (ii) the detection and analysis of binarity and transient phenomena; and (iii) the inference of atmospheric parameters.

\subsection{Automated Hot Subdwarf Candidate Selection in Large-Scale Surveys}

The primary challenge in hot subdwarf research is isolating these rare candidates from overwhelmingly large survey datasets. For photometric surveys, novel multiscale object detection algorithms, such as the Hot Subdwarf Detector (HsdDet), have been deployed to locate candidates directly within Sloan Digital Sky Survey (SDSS) images~\citep{2025AJ....170...94Z}. This work has identified 263 new hot subdwarfs, among 29,695 candidates detected in 216,014 images, using a spectral analysis tool XTGRID, e.g.,~\citep{2012MNRAS.427.2180N}.
Additionally, advanced unsupervised learning and dimensionality reduction algorithms, such as UMAP, have been used to reduce high-dimensional Gaia BP/RP (XP) spectra coefficients into 2D similarity maps~\citep{2026A&A...708A..23A}. These maps reveal that stellar BP-RP color is the dominant parameter driving the segregation of these stars from main sequence contaminants.

Spectroscopic data processing has seen equally significant advancements. The Large Sky Area Multi-Object Fiber Spectroscopic Telescope (LAMOST) data has been extensively exploited using Convolutional Neural Networks (CNNs), Support Vector Machines (SVMs), and Hierarchical Extreme Learning Machines (HELM) to detect hot subdwarfs~\citep{2017ApJS..233....2B,2019ApJ...886..128B,2019PASJ...71...41L,2022ApJS..259....5T,2024NatSR..1416815T}. These deep learning frameworks exhibit robust classification performance; for example, CNN-based hybrid models applied to LAMOST spectra have achieved testing accuracies of 96.17\%~\citep{2022ApJS..259....5T}. Similarly, kernel SVMs using Asymmetric Least Squares (ALS) baseline corrections achieved 87.0\% accuracy by identifying unique spectral patterns~\citep{2024NatSR..1416815T}. These ML pipelines systematically expand the known populations of both He-rich and He-poor hot subdwarfs.

\subsection{Detection of Binarity, Variability, and Transient Phenomena}
\label{sec3.2}

Because most models suggest a common-envelope binary evolution scenario for hot subdwarfs, detecting binarity and transient phenomena (e.g., eclipses, stellar flares) is a high priority. ML approaches using Self-Organizing Maps (SOMs) and CNN ensembles trained on normalized Gaia XP spectra have provided profound statistical insights into binary fractions~\citep{2024A&A...691A.223V}. Recent models predict that the binary ratio is approximately 18\% among hot subdwarfs showing no astrometric or photometric variability~\citep{2026A&A...708A..23A}. However, this fraction rises dramatically to over 60\% for targets with pure photometric variability, approximately 77\% for targets exhibiting both astrometric and photometric variability, and peaks at roughly 82\% for stars showing only astrometric variability.

Furthermore, ML algorithms leveraging summary statistics of sparsely sampled Gaia multi-band photometry (particularly the G-band amplitude) have successfully clustered distinct variable classes~\citep{2025A&A...693A.268R,2025A&A...704A..70R}, yielding over 89 new variable hot subdwarfs from combined Gaia and TESS observations. Most of these newly identified variables consist of reflection-effect and HW Vir-type binary systems. This approach efficiently separates subdwarfs from CVs without requiring expensive follow-up spectroscopy, providing an efficient way to assemble large, homogeneously selected samples of pulsating and non-pulsating hot subdwarfs. Such samples can, in turn, help empirically map the observed boundaries of the pulsation instability strip, keeping in mind that the instability strip itself is a theoretical construct and that the detection probability of pulsations, most of which are exhibited by sdB stars, can vary across the strip depending on pulsation amplitude and survey cadence.

\begin{figure}[H]
\includegraphics[width=0.52\linewidth]{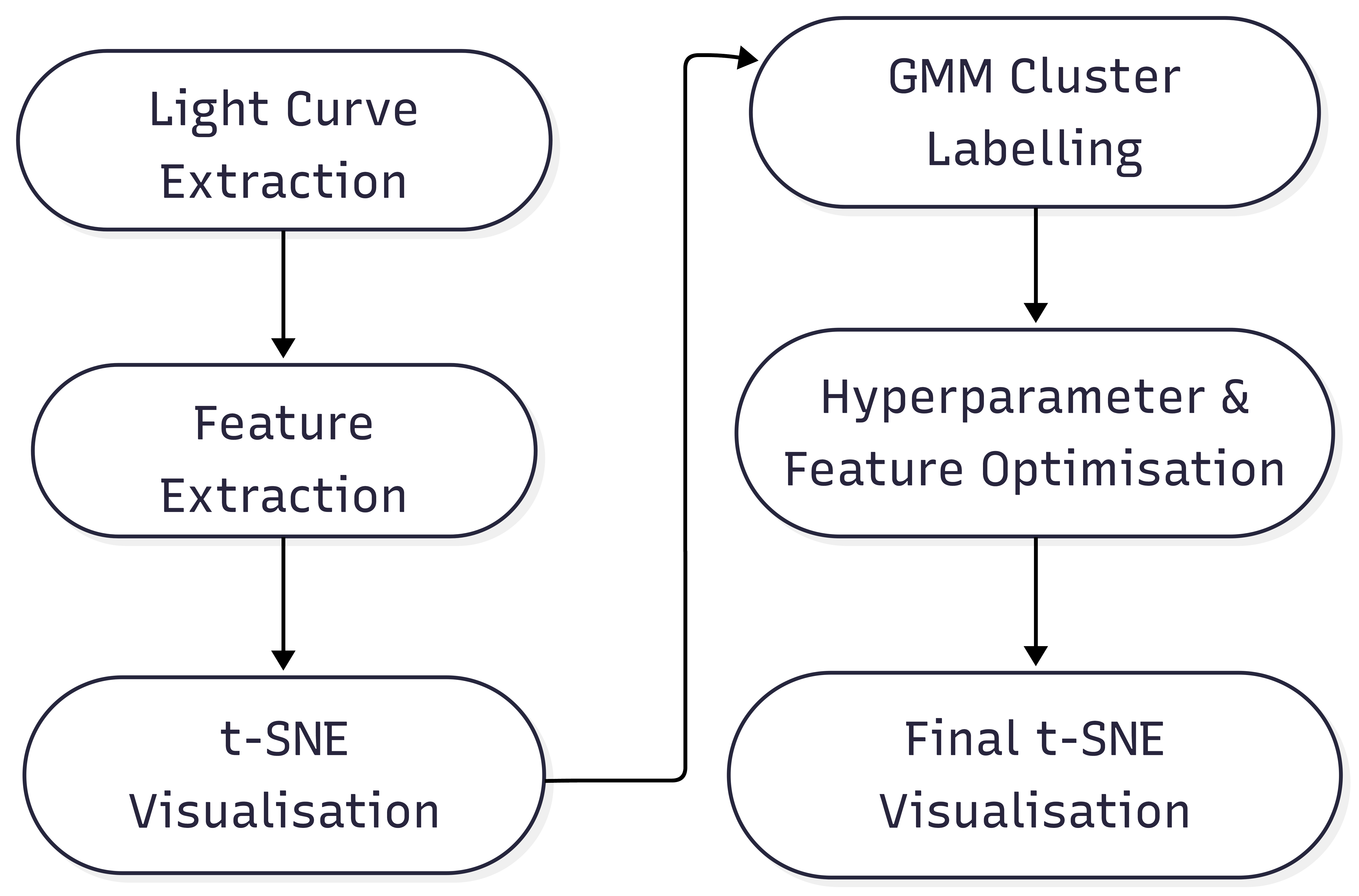}
\includegraphics[width=0.43\linewidth]{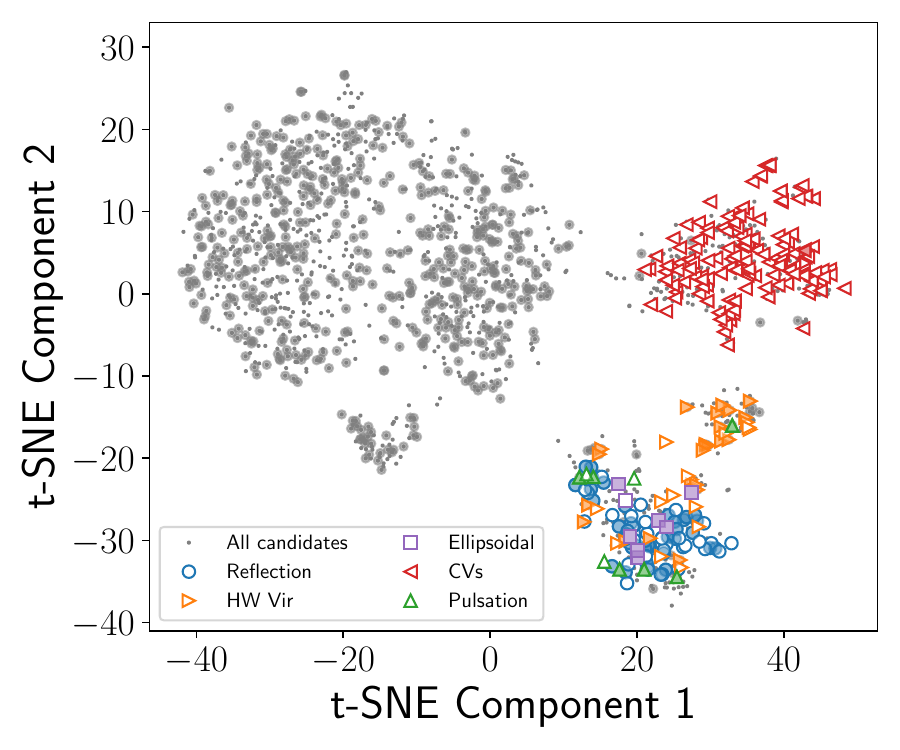}
\caption{(\textbf{Left}) panel:
Overview of the unsupervised machine learning pipeline applied to Gaia DR3 light curves of hot subdwarf candidates. (\textbf{Right}) panel: Resulting t-SNE embeddings (figure taken from Ranaivomanana et al.~\cite{2025A&A...693A.268R}), showing three main clusters that reflect three primary variability types. The three labeled clusters represent reflection-effect binaries, pulsators, and sources with dubious variability; the separation between clusters reflects primarily the light-curve morphology and amplitude used as input features, and the physical class assignment of individual sources within each cluster was subsequently verified against the literature classifications where available, rather than inferred from the embedding alone.}
\label{fig:tsne}
\end{figure}

\textls[-15]{This unsupervised clustering approach has been applied specifically to cluster Gaia DR3 light curves of confirmed and candidate hot subdwarfs~\citep{2025A&A...693A.268R}. Figure~\ref{fig:tsne} presents an overview of the corresponding ML pipeline applied to these data. The flowchart on the left panel of Figure~\ref{fig:tsne} summarizes the unsupervised learning steps. Approximately 1500 Gaia light curves were extracted, followed by a feature extraction phase that yielded 49 statistical and physical parameters after the removal of highly correlated ones. An initial t-SNE visualization was produced to explore the structure of the resulting feature space, followed by Gaussian Mixture Model (GMM) clustering to assign preliminary labels to the identified clusters. The pipeline then underwent iterative hyperparameter optimization, guided by the silhouette score and visual inspection of the graphical output. The silhouette score is a standard clustering-quality metric, ranging from $-1$ to $+1$, that measures how similar each point is to the other members of its own cluster relative to the nearest neighboring cluster; values close to $+1$ indicate compact, well-separated clusters, while values near zero or negative indicate overlapping or poorly defined clusters. Uninformative features were progressively removed using Random Forest feature importance ranking to further improve the clustering visualization. The optimal feature subset, which yielded the best silhouette score, was then used to generate the final t-SNE visualization shown alongside the flowchart.}

In addition to baseline variability, ML has been pivotal in identifying high-frequency transient events. Massive light-curve surveys from the Transiting Exoplanet Survey Satellite (TESS) and Evryscope have been leveraged using random forests, XGBoost, and neural networks to identify brief transit signals and flares~\citep{2020ApJ...890..126R,2024ApJS..271...57X}. Analyses of flares associated with compact stars strongly suggest that these events likely originate from unresolved cool main-sequence companions rather than the hot subdwarfs themselves~\citep{2024AJ....168..234L}. This confirms that binarity and environmental density heavily dictate the observed evolutionary paths of these systems.  

\subsection{Atmospheric Parameter Inference}
Beyond classification, advanced deep learning algorithms were recently used to predict hot subdwarf atmospheric parameters: effective temperature, surface gravity, and helium abundance. For instance, a CNN model with channel and spatial attention layers has been designed specifically for this task~\citep{2026ApJS..283...16L}. The model was trained on both 11,396 synthetic spectra and 945 observed spectra from LAMOST, achieving comparable prediction performance to traditional methods, with mean absolute errors of 730 K, 0.09 dex, and 0.03 dex in effective temperature, surface gravity, and helium abundance predictions, respectively. A similar CNN-based model had previously been developed for hot subdwarf identification and atmospheric parameter prediction~\citep{2024ApJS..274....2C}, though its prediction performance was slightly lower than that of the recent work. Despite this successful milestone in predicting physical parameters, the scarcity of observed spectra for these stars remains a hindrance to fully exploiting deep-learning-based prediction approaches, which require significantly large training samples.

\subsection{Outstanding Challenges Specific to Hot Subdwarf Classification}

Beyond the general contamination problem discussed above, several difficulties are specific to hot subdwarfs and deserve more explicit attention. First, the machine learning separability among hot subdwarf spectral subtypes (sdB, sdO, He-sdO, He-sdB, and sdOB) remains only partially explored: most existing classifiers are optimized to separate hot subdwarfs from contaminants rather than to resolve the finer subtype sequence, and the extent to which photometric or low-resolution spectroscopic features alone can recover this sequence is not yet well quantified. Second, an open question is whether classification outputs from ML pipelines can be used to indirectly constrain the dominant formation channel of a given hot subdwarf, such as common-envelope ejection, stable Roche-lobe overflow, or the merger of two helium white dwarfs. For instance, population-level correlations between inferred binary status, orbital period, and companion type may provide additional constraints on possible formation channels.. This remains largely unexplored territory for future work. Third, Gaia XP low-resolution spectra, while extremely valuable for large-scale candidate selection~\citep{2026A&A...708A..23A}, have intrinsic limitations in disentangling helium abundance, surface gravity, and contamination from a main-sequence companion, since these effects can produce degenerate signatures in the low-resolution BP/RP coefficients; higher-resolution spectroscopy remains necessary to break this degeneracy. Finally, distinguishing genuine HW~Vir-type eclipsing systems, reflection effects, ellipsoidal variations, and pulsations from CV contamination in TESS, Gaia, and ZTF light curves remains challenging when only sparse or single-band photometry is available, and misclassification between these variability types can bias derived binary fractions and instability-strip boundaries.

\section{Applications to White Dwarfs}
\label{methods_for_wds}

\textls[-10]{The rapid growth of large spectroscopic and photometric surveys has fundamentally transformed the study of white dwarfs, enabling population-scale analyses that were previously impractical. However, the volume, complexity, and heterogeneity of these data sets require the development of automated, data-driven methodologies for efficient classification, characterization, and discovery. In recent years, ML and advanced statistical techniques have been widely applied to white dwarf research, broadly falling into several key categories: (i)~automated classification and identification, (ii) discovery of rare and peculiar subclasses, (iii) analysis of binary systems, (iv) spectral modeling and parameter inference, and (v)~exploration of planetary debris and chemical composition. These categories can equivalently be viewed as addressing a sequence of scientific questions that recurs throughout this section: how the overall white dwarf sample is first assembled from Gaia HR-diagram selection and confirmed with SDSS/LAMOST spectroscopy (Section~\ref{wd_classification}); how confirmed white dwarfs are subsequently sorted into spectral classes such as DA, DB, DQ, DZ, DC, magnetic, and ultracool white dwarfs (Sections~\ref{wd_classification} and~\ref{wd_unsupervised}); how physical parameters such as effective temperature, surface gravity, mass, cooling age, and atmospheric composition are then estimated (Section~\ref{wd_spectral}); how special physical processes, including magnetic fields, metal pollution, debris disks, and binary interaction, are identified (Sections~\ref{wd_binary} and~\ref{wd_debris}); and how time-domain phenomena such as pulsation, rotational modulation, eclipsing binarity, and outburst or accretion events are detected. We follow this thematic thread within the subsections below, while noting that a dedicated systematic treatment of time-domain white dwarf phenomena, comparable to Section~\ref{methods_for_sdbs} (binarity, variability, and transient phenomena) for hot subdwarfs, is still comparatively underdeveloped in the literature and represents a natural direction for future dedicated reviews.}

Selection effects also need to be kept in mind when interpreting the results summarized below. Gaia color-absolute magnitude selection does not affect all white dwarf subpopulations equally.
\textls[-15]{Specifically, the selection tends to be least complete for the coolest, faintest white dwarfs; for composite-spectrum white dwarf--main-sequence (WD--MS) binaries whose photometry is diluted by the companion; for strongly magnetic white dwarfs whose broadband colors can be shifted by Zeeman splitting; and for lightly polluted white dwarfs whose metal features are only detectable at higher spectral resolution. Moreover, because many of the ML models discussed below are trained on spectroscopically confirmed subsamples from SDSS or LAMOST, their training data may not be representative of the full population of photometrically selected Gaia white dwarf candidates, so the derived class fractions and occurrence rates should be interpreted as conditional on the specific selection function of the training set rather than as unbiased estimates for the underlying population.}

\subsection{Automated Classification and Identification}
\label{wd_classification}

Efficiently classifying white dwarf spectra is a primary challenge for modern wide-field surveys, as the sheer volume of data has made manual visual inspection impossible. To address this, several ML pipelines have been developed to automate the process. For instance, models that integrate Gaia astrometry with SDSS spectroscopy achieve over 90\% accuracy, enabling the creation of catalogs containing hundreds of thousands of objects~\citep{2023MNRAS.521..760V}. More advanced convolutional neural networks, in particular multimodal versions that process both 1D spectra and 2D images, have pushed classification accuracy toward 99\%, significantly reducing the requirement for human intervention~\citep{2025ApJS..279...36Z}.
In addition to spectral analysis, new object-detection frameworks can now identify white dwarf candidates directly from imaging data. By bypassing traditional photometric color-cut limitations, these methods have successfully uncovered tens of thousands of new candidates within massive survey datasets~\citep{2025ApJS..276...53Z}.

\subsection{Unsupervised Learning and Discovery of Rare Subclasses}
\label{wd_unsupervised}

Unsupervised ML techniques have proven particularly powerful for identifying rare or previously unknown subclasses of white dwarfs. Dimensionality reduction methods such as UMAP and t-SNE have been widely used to map high-dimensional spectral data into lower-dimensional spaces, revealing distinct clusters corresponding to different spectral types and physical properties~\citep{2024MNRAS.535.2246B,2025RASTI...4af044B,2025A&A...704A..70R,2026MNRAS.546ag246M}. These methods have enabled the discovery of new polluted white dwarf candidates, the identification of magnetic white dwarfs, and the detection of rare spectral features that may be missed by supervised classification. For example, clustering techniques combined with density-based algorithms have successfully separated magnetic and non-magnetic populations and enabled the estimation of magnetic field strengths in cases where direct measurements are unavailable~\citep{2026arXiv260311945K}. Similarly, self-organizing maps and other unsupervised approaches have identified new metal-polluted white dwarfs in Gaia data, demonstrating their effectiveness in mining large, sparsely labeled datasets~\citep{2024ApJ...977...31P,2024ApJ...970..181K}.

\subsection{Binary Systems and Population Studies}
\label{wd_binary}

\textls[-15]{Data-driven approaches have also been widely applied to the identification and characterization of white dwarf binary systems, particularly WD--MS binaries. Neural network-based approaches combined with Gaussian process classification have enabled the identification of tens of thousands of WD--MS candidates in Gaia data, including systems in which the main-sequence companion dominates the observed spectrum~\citep{2025ApJS..279...47L}. In addition, classification techniques have been extended to simulated datasets to prepare for future observations. For example, ML models trained on synthetic spectra have demonstrated strong performance in classifying WD--MS systems in low-resolution Gaia spectra~\citep{2022A&A...667A.144E}. Beyond electromagnetic observations, ML is also being applied to gravitational-wave astronomy, where it is used to identify compact binaries detectable by future missions such as LISA, including double white dwarf systems and rarer neutron star--white dwarf binaries~\citep{2026arXiv260306341T}.}

\subsection{Spectral Modeling and Parameter Inference}
\label{wd_spectral}

Another major application of ML in white dwarf studies is the inference of stellar parameters and chemical abundances from spectroscopic data. Traditional spectral fitting techniques are computationally expensive and difficult to scale to large datasets. To address this, ML-based spectral modeling frameworks have been developed to generate synthetic spectra and perform parameter estimation. For example, neural-network-based models trained on large grids of atmosphere models can infer effective temperature, surface gravity, and abundances with accuracies comparable to classical methods~\citep{2024MNRAS.529.1688B}. These approaches have also been successfully applied to detailed abundance analyses of metal-polluted white dwarfs, enabling efficient characterization of multiple chemical species from spectroscopic observations, e.g.,~\citep{2025MNRAS.540..746B}.

\subsection{Planetary Debris and Chemical Composition}
\label{wd_debris}

White dwarfs polluted by planetary debris provide a unique opportunity to study the composition of exoplanetary material. ML techniques have significantly improved the identification and characterization of these systems. Unsupervised clustering methods applied to Gaia spectra have revealed distinct groups of metal-polluted white dwarf candidates and increased the number of known systems after confirming their physical nature through spectroscopic follow-up observations~\citep{2024ApJ...970..181K}. Comparative studies of different ML approaches have shown that both supervised and unsupervised methods play complementary roles in identifying polluted white dwarfs. While supervised models benefit from labeled training data, unsupervised techniques are particularly effective at discovering new candidates and rare subclasses~\citep{2025RASTI...4af044B}. These developments are enabling large-scale studies of planetary system evolution and the chemical diversity of exoplanetary material.
 
\section{Conclusion and Future Prospects}
\label{conclusion}

In this review, we have summarized how the interplay between large-scale data from modern astronomical facilities and advanced ML techniques is reshaping the identification and characterization of hot subdwarfs and white dwarfs. Various ML models have been implemented in stellar astronomy, ranging from classical algorithms to deep learning architectures, using both supervised and unsupervised approaches. We have also seen that astronomical facilities generate a wide variety of data, including tabular catalogs, images, spectra, and time series, at rates reaching up to 15 TB per night (as an integrated nightly science product; see Table~\ref{tab:data_rates_facilities}). Consequently, various ML tasks emerge, such as object classification, transit detection, and astrophysical parameter estimation.

Table~\ref{tab:sdb_wd_comparison} summarizes, side by side, the main similarities and differences between hot subdwarfs and white dwarfs that motivate their joint treatment in this review, together with the machine learning tasks and open challenges specific to each class.

\begin{table}[H]
\caption{Comparison of hot subdwarfs and white dwarfs as case studies for machine-learning-driven discovery in late stellar evolution.}
\label{tab:sdb_wd_comparison}
\begin{adjustwidth}{-\extralength}{0cm}
\begin{tabularx}{\fulllength}{lLL}
\toprule
\textbf{Aspect} & \textbf{Hot Subdwarfs} & \textbf{White Dwarfs} \\
\midrule
Evolutionary stage & 
Core-helium-burning, late but not final &
Final compact remnant \\

Dominant formation channel &
Binary mass loss (CE, RLOF, white dwarf merger) &
Single- and binary-star envelope loss \\

Main contaminants &
BHB stars, CVs, low-mass main-sequence stars, white dwarfs &
Faint main-sequence/subdwarf stars, WD--MS composites \\

Dominant ML tasks &
Candidate selection, variability/binary classification &
Spectral classification, parameter and abundance inference \\

Main open ML challenge &
Subtype separability (sdB/sdO/He-sdB/He-sdO/sdOB); scarce labeled spectra &
Selection-function bias from spectroscopic training sets; rare-subclass discovery \\
\bottomrule
\end{tabularx}
\end{adjustwidth}
\end{table}

In hot subdwarf research, ML techniques are still largely applied to candidate identification and variability classification, while little work has been done regarding physical parameter estimation, such as predicting atmospheric parameters and abundances. This limitation is mainly due to the scarcity of labeled data and the high cost of acquiring the spectral data needed to confirm genuine candidates. Nevertheless, the integration of unsupervised ML methods has significantly improved the detection of hot subdwarf candidates, enabling robust statistical studies and physical inferences about these stars. In particular, unsupervised ML approaches have been shown to efficiently categorize large datasets of hot subdwarfs and contaminants using photometric data alone, which were traditionally difficult to separate without additional spectroscopy.

The application of ML techniques and advanced statistical methods has similarly reshaped white dwarf research across multiple fronts. Automated classification pipelines now enable the processing of vast spectroscopic and photometric surveys while maintaining consistent accuracy at scale, though, as discussed in Section~\ref{methods_for_wds}, this accuracy is conditioned on the completeness and representativeness of the underlying training samples. Unsupervised learning techniques have been particularly effective in revealing rare and previously unknown subclasses, including magnetic and metal-polluted white dwarfs, by uncovering hidden structure in high-dimensional survey data. In parallel, ML techniques have improved the identification and characterization of binary systems, supported scalable spectral modeling and parameter inference, and enabled large-scale studies of chemically polluted atmospheres that trace the composition of exoplanetary material. Together, these developments demonstrate that data-driven approaches are becoming essential tools for extracting physical insight from the rapidly expanding white dwarf datasets when combined with the follow-up validation discussed above, and they will continue to play a central role in future survey-driven discoveries.

However, despite these substantial advances, important limitations remain. Specifically, the performance of current classification and inference methods remains sensitive to data quality. In low signal-to-noise regimes, where spectral features become ambiguous and photometric variability is poorly constrained, current ML models can produce incomplete or uncertain classifications. This reveals gaps in the parameter space where certain classes of objects are systematically underrepresented or misidentified due to selection effects.

To address these issues, future research might explore combining a data-centric approach with ML architecture improvements. That is, rather than focusing solely on optimizing ML models, the community would benefit from actively curating higher-quality training sets. This can be done, for instance, through active learning strategies that prioritize targeted spectroscopic follow-up of rare or ambiguous sources~\citep{2019MNRAS.483....2I}, synthetic data generation for underrepresented classes~\citep{2022ApJ...935...23G}, or
multimodal learning approaches that combine heterogeneous data via cross-survey cross-matching to mitigate selection biases~\citep{2026arXiv260700228S,2026arXiv260518959G}. At the same time, model-side advances remain essential; in particular, uncertainty-aware methods such as Bayesian inference and conformal prediction~\citep{2020MNRAS.491.1554W,2026arXiv260705393B}, as well as physics-informed architectures~\citep{2024arXiv240300599B,2024ApJ...975..258A} that incorporate domain knowledge as structural constraints, offer complementary paths toward more interpretable and well-calibrated classifications.

More concretely, we identify the following routes as particularly promising over the coming years: (i) the establishment of public, community-agreed benchmark datasets for hot subdwarfs and white dwarfs, with standardized train/validation/test splits and documented selection functions, to enable direct comparison between competing ML methods; (ii) a systematic shift toward calibrated probabilistic outputs, rather than hard class labels, so that downstream population studies can propagate classification uncertainty; (iii) the incorporation of stellar atmosphere models, evolutionary tracks, and color-magnitude diagram priors, together with observational uncertainties, directly into the loss functions of physics-informed architectures, so that model outputs remain consistent with known stellar physics; and (iv) the design of low-latency, real-time classification pipelines capable of triaging compact stellar objects directly within the Rubin/LSST alert stream, where the volume of nightly alerts will preclude purely manual vetting.

Beyond these technical and methodological improvements, it is worth emphasizing the fundamental astrophysical questions that ML-enabled samples may help address as survey volumes continue to grow. As Rubin/LSST, WEAVE, 4MOST, and future Gaia-like missions deliver homogeneously selected samples of hot subdwarfs and white dwarfs that are orders of magnitude larger than those available today, ML-based approaches could enable population-level physical inference from large, homogeneous samples of compact stars. For example, sufficiently large and well-characterized samples of binary hot subdwarfs, classified and parameterized at scale, could provide new empirical constraints on the common-envelope efficiency parameter and on the relative importance of different binary mass-loss channels, which are currently poorly constrained by the comparatively small samples available from targeted spectroscopic follow-up~\citep{2002MNRAS.336..449H,2003MNRAS.341..669H,2010A&A...520A..86Z,2022ApJ...933..137G}. Similarly, large, homogeneously selected samples of pulsating compact stars could refine the empirical boundaries of the instability strips and improve understanding of pulsation-mode selection of the hot subdwarf and white dwarf instability strips, complementing and testing current theoretical pulsation-driving models~\citep{2003ApJ...583L..31G,2010A&A...513A...6O,2010A&ARv..18..471A,2024A&A...684A.118U}. For white dwarfs, statistically robust, ML-derived samples of cooling ages, masses, and atmospheric compositions could sharpen the empirical initial--final mass relation~\citep{2010A&ARv..18..471A,2015MNRAS.450.3708R,2018ApJ...866...21C,2018ApJ...860L..17E} and allow white dwarf cooling sequences to be used as reliable chronometers for Galactic archaeology on a much larger scale than is currently possible~\citep{2015arXiv150602653G,2019MNRAS.482..965K}. \textls[-15]{Larger samples of metal-polluted white dwarfs, efficiently identified through ML pipelines, could likewise enable population-level statistical studies of the bulk composition of exoplanetary material~\citep{2016NewAR..71....9F,2021MNRAS.504.2853H}, opening the possibility of investigating whether the composition of planetary material varies with Galactic environment or progenitor stellar properties. Realizing this potential, however, will require the methodological advances discussed above, in particular robust uncertainty quantification and careful control of selection effects, since population-level physical inferences drawn from ML-classified samples are only as reliable as the completeness and purity of the underlying catalogs.}

Furthermore, degeneracies between different astrophysical scenarios can make the interpretation of the data more difficult. A notable example is the confusion between short-period binary systems and long-period pulsating hot subdwarfs or white dwarfs, which can produce similar photometric signatures in single-epoch or sparsely sampled data, as demonstrated in~\citet{2025A&A...693A.268R,2025A&A...704A..70R}. These challenges underscore the critical importance of dedicated follow-up observations. Multi-epoch spectroscopy and time-series photometry are essential to confirm candidate classifications, break parameter degeneracies, and accurately disentangle variability mechanisms. Such observations provide the temporal baseline required to distinguish orbital motion from pulsational behavior and to improve parameter constraints derived from ML models. Therefore, while ML methods are indispensable for efficiently mining large surveys, their full scientific potential can only be realized when combined with systematic follow-up campaigns that ensure robust validation and physical interpretation of the identified sources.

\authorcontributions{\textls[-15]{Conceptualization, P.R. and M.U.; methodology, P.R. and M.U.; \mbox{writing---original} draft preparation, P.R. and M.U.; writing---Sections~\ref{methods}~and~\ref{methods_for_sdbs}, P.R.; writing---abstract, Sections~\ref{intro}~and~\ref{methods_for_wds}, M.U.; writing---Section~\ref{conclusion}, 
P.R. and M.U.; visualization, P.R.; software, P.R.; investigation, P.R. and M.U.; writing---review and editing, P.R. and M.U.; project administration, P.R and M.U. All authors have read and agreed to the published version of the manuscript.}}

\funding{M.U. was supported by the Research Foundation Flanders (FWO) through a Postdoctoral Fellowship (grant agreement No. 1247624N). P.R. received no specific external funding for this work.}
 
\dataavailability{No new data were generated in this review. All data discussed and summarized in this manuscript are publicly available in the original publications cited throughout the text and in the references therein. The Gaia DR3 data used to produce Figure~\ref{fig:hrd} are publicly available through the Gaia Archive (\url{https://gea.esac.esa.int/archive/}).}

\acknowledgments{The authors acknowledge the use of Mermaid Live Editor (https://mermaid.live), accessed on 15 June 2026, for the purposes of creating the diagram in the left panel of Figure 2.}

\conflictsofinterest{The authors declare no conflicts of interest.}

\begin{adjustwidth}{-\extralength}{0cm}

\reftitle{References}



\PublishersNote{}
\end{adjustwidth}
\end{document}